\documentclass[a4paper,fleqn,11pt]{article}
\usepackage{graphicx,array}
\usepackage{amssymb,amsmath}
\usepackage{cite}
\usepackage{float}
\usepackage{comment}

\renewcommand{\normalsize}{\fontsize{9pt}{15pt}\selectfont}
\newcommand{\halfnormalsize}{\fontsize{9pt}{12pt}\selectfont}
\renewcommand\title[1]{
\begin{center}\fontsize{12pt}{15pt}{\bf{\bfseries\sffamily#1}\par}\end{center}}
\renewcommand\author[1]{\begin{center}\small{\sffamily#1}\end{center}\par}
\renewcommand\abstract[1]{\begin{quote}\footnotesize #1\end{quote}\par}
\makeatletter
\renewcommand{\fnum@figure}[1]{\fontsize{8pt}{12pt}{\textbf{Figure~\thefigure. }}}
\renewcommand{\fnum@table}[1]{\fontsize{8pt}{12pt}{\textbf{Table~\thetable. }}}
\@ifundefined{captionstyle}{}{}
\@ifundefined{instindent}{\newdimen\instindent}{}
\long\def\@caption#1[#2]#3{\par\addcontentsline{\csname
  ext@#1\endcsname}{#1}{\protect\numberline{\csname
  the#1\endcsname}{\ignorespaces #2}}\begingroup
    \@parboxrestore\if@minipage\@setminipage\fi
    \@makecaption{\csname fnum@#1\endcsname}{\ignorespaces #3}\par
  \endgroup}
\renewcommand\@seccntformat[1]{\csname prefmt@#1\endcsname
	\csname the#1\endcsname. \csname postfmt@#1\endcsname}
\newcommand\postfmt@section{\hskip 1mm}
\newcommand\postfmt@subsection{\hskip 1mm}
\renewcommand\section{\@startsection{section}{1}{\z@}%
{-12pt plus3pt minus2pt}%
{3pt}%
	{\normalfont\normalsize\bfseries}}
\renewcommand\subsection{\@startsection{subsection}{2}{\z@}%
{-9pt plus3pt minus2pt}%
{3pt}%
	{\normalfont\normalsize\bfseries}}
\renewcommand\subsubsection{\@startsection{subsubsection}{3}{\z@}%
{-9pt plus3pt minus2pt}%
{3pt}%
	{\normalfont\normalsize\bfseries}}
\makeatother
\newcommand\stamp[6]{\vspace{0.25cm}\linespread{1}\noindent\footnotesize
#1,\ #2:\ #3,\ #4 ({\em #5}).\ #6\par}
\def\R{\mathbb R}

\def\({\left(}
\def\){\right)}

\usepackage{amsmath,amsfonts,units}

\newcommand{\field}[1]{\mathbb{#1}}

\begin{document}
\vspace*{5.4 mm}
%
%

\title{A bifurcation and symmetry discussion of the Sommerfeld effect}
\vspace*{3.8 mm}
%
%
\author{Eoin Clerkin, Rubens Sampaio}
%
%

\abstract{{\em Abstract:} 

The Sommerfeld effect is an intriguing resonance capture and release series of events originally demonstrated in 1902.
A single event is studied using a two degree of freedom mathematical model of a motor with imbalance mounted to laterally restricted spring connected cart.
For a certain power supplied, in general the motor rotates at a speed consistent with a motor on a rigid base.
However at speeds close to the natural frequency of the cart, it seemingly takes on extra oscillations where for a single rotation it both speeds up and then slows down.
Therefore in a standard experimental demonstration of the effect, as the supplied torque force is increased or decreased, this may give the illusion that the stable operation of the motor is losing and gaining stability.
This is not strictly the case, instead small oscillations always present in the system solution are amplified near the resonant frequency.
The imbalance in the motor causes a single resonance curve to fold back on itself forming two fold bifurcations which leads to hysteresis and an asymmetry between increasing and decreasing the motor speed.
Although the basic mechanism is due to the interplay between two stable and one unstable limit cycles, a more complicated bifurcation scenario is observed for higher imbalances in the motor.
The presence of a $\field{Z}_2$ phase space symmetry tempers the dynamics and bifurcation picture.
}

\section{Historical background of and introduction to the Sommerfeld effect}\label{sec:intro}

Arnold Sommerfeld's posthumous biography\cite{Arnold_Sommerfeld} complied from his handwritten correspondence, 
mentions an intriguing engineering problem proposed by Prof.\ Hermann Boost also of the Technical University of Aachen
whereby a steam engine is to be installed in a building which is itself to be supported by beams.
As a demonstration to the district association of German engineers (VDI), 
using a weighted motor with small imbalance screwed tightly to a tabletop,
he caused the motor to become enthralled to the frequency of the table
which induced large vibrations highlighting to the audience the catastrophic potential of resonance.
The phenomenon was explained 
with reference to resonance curves for harmonic-forced linear differential equations 
in his seminal 1902 paper~\cite{Sommerfeld1902}.
It was fifty years after with the work of Blekhman~\cite{Blekhman1953,Fidlin2005} that the effect was named in his honour 
and more than sixty years later it has seen a renewed interest among the applied mechanics and mechanical engineering community, 
being a major discussion point at the 2015 IUTAM symposium in Frankfurt~\cite{Dantas2016,Fidlin2016,Hagedorn2016}. \\


\par

\noindent\begin{minipage}{0.45\textwidth} 
\begin{figure}[H]
\includegraphics[width=1.0\linewidth]{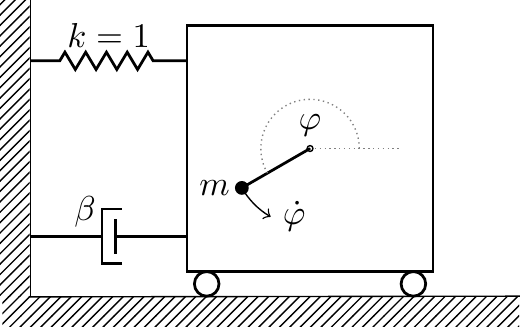}
\caption{\label{fg:scheme} Schematic diagram of laterally restricted spring-connected and damped cart with a driven rotor with imbalance.}
\end{figure}
\end{minipage}\hfill\begin{minipage}{0.425\textwidth} 
\begin{figure}[H]
\includegraphics[width=1.0\linewidth]{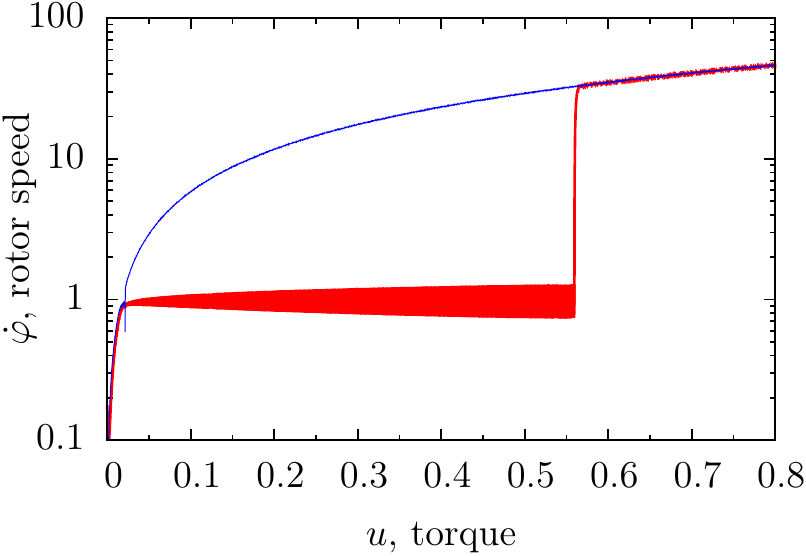}
\caption{\label{fg:sweep} Sweep diagram showing the rotor speed as the torque force is slowly increased (red) and decreased (blue).}
\end{figure}
\end{minipage}\\[1em]

In order to study the Sommerfeld effect as a purely mechanical phenomenon, 
a translational oscillator rotational actuator, as schematically drawn in Fig.~\ref{fg:scheme},
has been investigated by a number of authors~\cite{Fradkov2011,Dantas2016,Fidlin2016,Kiseleva2016}
as a minimal model that is believed to encompasses only its essential dynamical attributes.
This paradigmatic example is made up of a rotor with a small imbalance (m) a certain distance from its centre which is mounted to a laterally-restricted spring-connected and damped cart.
This drawing (Fig.~\ref{fg:scheme}) follows the normal applied mechanics convention whereby a single wall span implies restriction in two spatial directions.
In addition, the rotor would normally be orientated in the horizontal plane, so as to remove any need to consider gravitational effects.
Physical parameters expected to be important to model such a scenario would be mass inertia of the cart, imbalance and rotor, the spring and damping constants as well as a measure of the symmetry-brokenness in the rotor, i.e the distance between the imbalance and its centre of rotation.
The supplied power or torque force to the rotor is varied to unveil the Sommerfeld effect.
\noindent\begin{figure}[t]
\center
\includegraphics[width=0.8\linewidth]{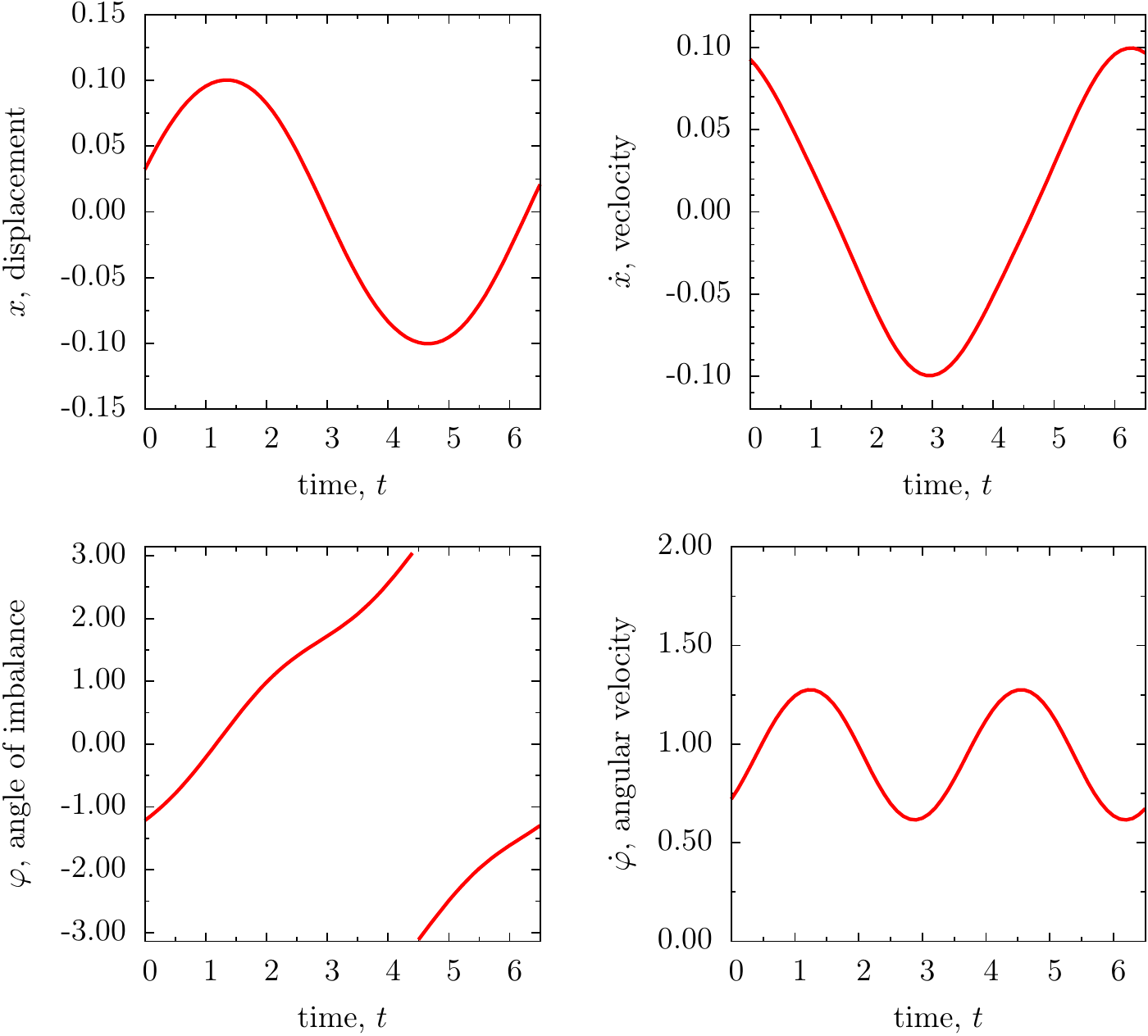}
\caption{\label{fg:symm} For $u=0.1$ and $\epsilon=0.0125$, time-traces showing a symmetric, resonance captured S-type 
cycle~\eqref{eq:slc}. Period $\tau=\frac{2\pi}{avg(\dot{\varphi})}=\frac{6.28}{0.94}=6.68$. 
$x$ can be seen to precede $\dot{x}$ by $\frac{\tau}{4}$.}
\end{figure}

\par

As shown by the red curve in Fig.~\ref{fg:sweep}, as the power is monotony increased to the non-ideal~\cite{Evan1976} motor,
its angular velocity increases until it approaches the natural frequency of the spring,
which is normalised to one in this manuscript.
After which, additional increases in power does not increase the motor speed, 
but instead leads to the growth of oscillations in the smooth operation of the motor and in the displacement of the cart (not seen).
A log-scale diagram of rotor speed versus the supplied torque is shown in Fig.~\ref{fg:sweep} where a linear response of the angular velocity to the torque appears as a logarithm function. 
Subsequently, as the torque force is increased further (red curve), 
there comes a point where the oscillations in the cart fall relatively silent 
and the rotor rapidly speeds up to match the expected speed of the rotor had it been on a rigid base.
Characteristic of the Sommerfeld effect is an asymmetry between increasing (red curve) and decreasing (blue curve) the power to the motor,  and significant hysteresis is seen, meaning that two distinct states for the system are concurrently stable.
In fact, this was alluded to in the 1902 paper~\cite{Sommerfeld1902} where Sommerfeld discussed temporarily grasping the table legs to change the motors speed to a state operating at higher frequency.
In our case, when the system is operating in the resonance captured zone, a temporarily restriction on cart's movement would allow the rotor to be released from resonance.
The succeeding section to the next will explain the dynamics behind Fig.2, 
but first we introduce the equations of motion which were integrated to generate it.

\section{Mathematical model of a translational oscillator rotational actuator}\label{sec:basic}

In order to investigate a single Sommerfeld resonance capture and release event, a model for a translational oscillator rotational actuator as schematically drawn in Fig.~\ref{fg:scheme} is studied.
The kinetic and potential energies of the cart and rotor are available from Appendix B of Ref.~\cite{rand1992} which can be used to derive the system of equations with dimensions such as those in Refs.~\cite{Kiseleva2016,Fradkov2011}.
The following work uses the non-dimensionalised version of these equations from Ref.~\cite{Fidlin2005} which significantly reduces the number of required parameters to consider.
\begin{subequations}\label{eq:orig}
\begin{equation}\label{eq:orig:a} d_t^2 {x} + \beta d_t {x} + x = - \epsilon d_t^2\left(\cos{(\varphi)}\right),  \end{equation}
\begin{equation}\label{eq:orig:b} d_t^2 \varphi + \nu d_t \varphi = u + \epsilon \kappa \sin(\varphi) d^2_t {x}.\end{equation}
\end{subequations}
The system has two degrees of freedom, 
the displacement of the cart $x$, 
defined positively to the right and negatively to the left from its equilibrium position, 
and $\varphi$ as the angle of the imbalance in the rotor, defined by the normal mathematics convention from the right horizontal axis as shown in Fig.~\ref{fg:scheme}. 
In this paper $d_t$ to the left of or a dot above a variable represents the operation of differentiating with respect to non-dimensionalised time~$t$.

\par

The non-dimensionalised parameter $\beta, \epsilon, \kappa, \nu$ encapsulates dependant dimensioned parameters such as the mass of the cart and imbalance, damping in the cart and rotor, level of symmetry-brokenness, and moment of inertia.
Parameter values were chosen to compare with the work in Refs.~\cite{Kiseleva2016,Fradkov2011}.
After transforming these parameter to their non-dimensionalised form, the parameter values become $\beta=0.01983$, $\kappa=0.017$, $\nu=857.143$ and the natural frequency in the spring-cart is normalised to one (cf.\ Fig.~\ref{fg:sweep}).
$\epsilon=\{0.005, 0.0125\}$ is one of the more important parameter as a measure of imbalance or symmetry-brokenness in the rotor which results in the coupling between relations~\eqref{eq:orig:a} and~\eqref{eq:orig:b} in Eq.~\eqref{eq:orig}.
For this parameter, the following symmetry exists.
\begin{equation}\label{eq:x_eps}\begin{bmatrix}x \\ \epsilon\end{bmatrix} \longmapsto -\begin{bmatrix} x\\ \epsilon \end{bmatrix} \qquad\qquad\qquad\qquad\qquad \text{parameter symmetry}\end{equation}
Physically $\epsilon$ is proportional to the length between the imbalance and its centre of rotation and therefore the parameter symmetry~\eqref{eq:x_eps} may be interpreted as a redefinition of orientation of $x$ displacement as the imbalance is translated by 180$^o$ after going to negative length.
The torque $u$ is the force supplied to the rotor, which is the parameter used to unveil the Sommerfeld effect and is the main sweep parameter in this study.
A \emph{parameter space symmetry} also exists, namely 
\begin{equation}\label{eq:slc}\begin{bmatrix}\varphi \\u\end{bmatrix}\longmapsto-\begin{bmatrix}\varphi\\u\end{bmatrix} \qquad\qquad\qquad\qquad\qquad \text{parameter symmetry}\end{equation}
which allows one to obtain the dynamics due to clockwise driving of the rotor from the anti-clockwise driving by means of the additive inverse of the displacement and angle, thereby velocity and angular velocity of the cart and rotor. 
Because of these two parameter symmetries, we may limit this study to positive torque and positive $\epsilon$ only and still obtain the full dynamical picture.

\par

Unlike the two introduced parameter systems, a \emph{phase space} symmetry such as the following involutionary (a.k.a reflection) symmetry influences the dynamics by itself.
\begin{equation}\label{eq:z2}\begin{bmatrix}x \\\varphi\end{bmatrix}\longmapsto\begin{bmatrix}- x \\\varphi + \pi \end{bmatrix} \qquad\qquad\qquad\qquad\qquad \field{Z}_2\text{ symmetry}\end{equation}
This means that if $y_1(t)$ is a solution to the Eq.~\eqref{eq:orig} then so is $y_2(t)=\mathbf{R} y_1(t)$ where $\mathbf{R}$ is the action of the symmetry.	
As this study's primarily interest is in periodic orbits in the system as the resonance frequency of the cart is transversed by the rotor, it is of importance to consider how symmetry~\eqref{eq:z2} may affect limit cycles. 
To this aim, we will use the results and language of Refs.~\cite{Nikolaev1995,Nikolaev1998a}.
The phase space may be decomposed $\R^4 = X^+\oplus X^-$ where the action of the symmetry~\eqref{eq:z2} is $R v = v \text{ for } v \in X^+ $ and $R v = -v \text{ for } v \in X^-$.
Some care needs be exercised when doing this as $\varphi$ is not strictly in $\R$, so we consider various coordinate transforms such as $\{\varphi,\dot{\varphi}\}=\{\dot{\varphi}\cos{\varphi},\dot{\varphi}\sin{\varphi}\} \in \R^2$ which enforces this.
It may be thus deduced that $X^+ =\emptyset$ and this has the immediate consequence to limit the types of limit cycles permitted in system~\eqref{eq:orig}, 
namely limit cycles of fixed or mixed-fixed symmetry type of Refs~\cite{Nikolaev1995,Nikolaev1998a} do not exist.
The only cycles remaining which are \emph{invariant} to $\field{Z}_2$ symmetry~\eqref{eq:z2} are of the following type
\begin{equation}\label{eq:slc}
\begin{bmatrix}
x \\
\varphi
\end{bmatrix}(t)
\longmapsto
\begin{bmatrix}
-x \\
\varphi + \pi
\end{bmatrix}(t+\frac{\tau}{2}), \qquad\qquad\qquad\qquad\qquad \text{S- and M-type}
\end{equation}
where $\tau$ is the minimal period of the limit cycle. 
In the nomenclature of Refs.~\cite{Nikolaev1995,Nikolaev1998a}, these are called symmetric or S-type cycles.
As can be seen after half the period of oscillation the displacement is exactly its negative.
Likewise this rule applies for the velocity.
Therefore, cycles invariant with respect to condition~\eqref{eq:slc} must average to zero for these variables over one cycle.
A typical limit cycle of system Eq.~\eqref{eq:orig} is shown in Fig.~\eqref{fg:symm} which can be seen to be invariant with respect to Eq.~\eqref{eq:slc}.
It is known~\cite{Klivc1986} that a period doubling bifurcation may not occur in limit cycles of S-type due to a multiplicity of two in their Floquet multipliers.
In later sections, cycles which are symmetry-broken with respect to Eq.~\eqref{eq:slc} will be discussed.

\section{Basic mechanism of the Sommerfeld effect}\label{sec:basic}

By direct integration of the equations of motion~\eqref{eq:orig} with different initial angular velocities, Refs.~\cite{Kiseleva2016,Dudkowski2016} highlight the existence of ''hidden'' bistable attractors.
The 2D diagram Fig.~\ref{fg:3lc} shows the ultimate limit cycles of these trajectories by the red and blue solid closed periodic orbits. 
The red curve shows large variance in the displacement and velocity, hence large potential and kinetic energy in the cart.
Conversely, although the blue curve shows higher eccentricity thus a higher ratio of kinetic to potential energy in the cart, 
both energies are substantially less than that of the red limit cycle, 
instead it has considerably greater total energy in the rotating rotor.
This can be seen in Fig.~\ref{fg:contin} where the blue limit cycle has larger maximum angular velocity than the red limit cycle.
In the case of the blue limit cycle, the rotor operates at the frequency consistent with a motor on a rigid base 
whilst in the case of the red limit cycle, the rotor rotates at an average frequency approximately consistent with the natural frequency of the cart.
This oscillation frequency persists over an extensive change of torque as can be seen in Figs.~\ref{fg:sweep} and~\ref{fg:contin} and the red limit cycle is captured by the resonance.
In the resonance captured range of torque, there is therefore at least bistability of limit cycle states. 

\noindent\begin{minipage}{0.4\textwidth} 
\begin{figure}[H]
\includegraphics[width=1.0\linewidth]{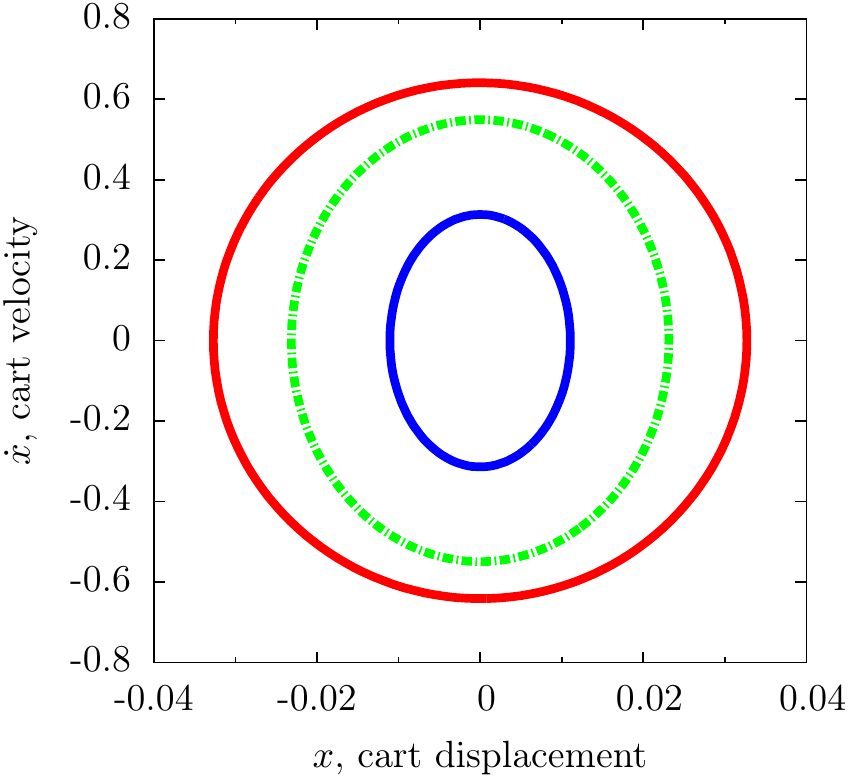}
\caption{\label{fg:3lc} Two stable limit cycles (red and blue) separated by an unstable limit cycle (green - dashed). Generated for $u=0.15$ and $\epsilon=0.005$.}
\end{figure}
\vspace{0.5em}
\end{minipage}\hspace{2.5em}\begin{minipage}{0.53\textwidth} 
\begin{figure}[H]
\includegraphics[width=1.0\linewidth]{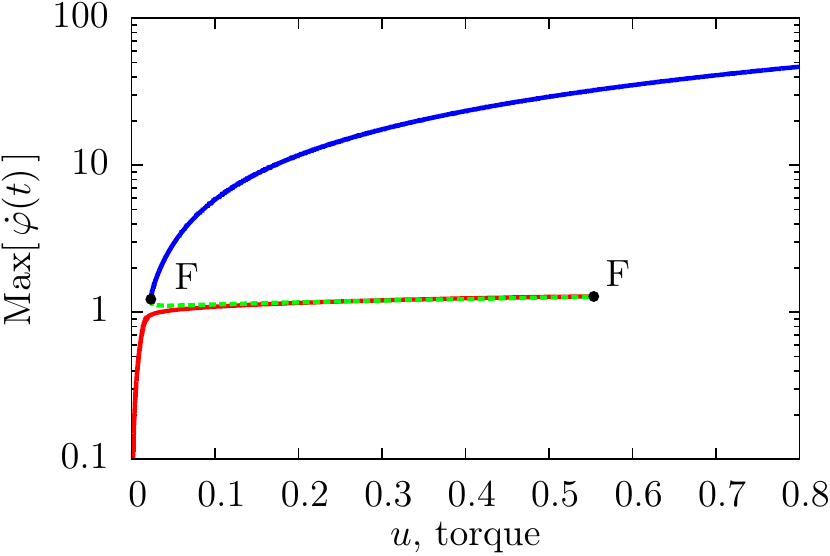}
\caption{\label{fg:contin} Continuation of the S-type cycle at $\epsilon=0.005$ using AUTO~\cite{AUTO} showing a loss and gain of stability at fold bifurcations of limit cycles, labelled by F. (cf.\ Fig.~\ref{fg:sweep})}
\end{figure}
\vspace{0.5em}
\end{minipage}

In fact, a third limit cycle exists between the resonance captured and resonance released dynamics as shown in Figs~\ref{fg:3lc} and~\ref{fg:contin} by the green dashed line.
This limit cycle has intermediate energy in the cart and although it is unstable, it controls the limits of the stable dynamics.
As the torque is increased from its value in Fig.~\ref{fg:3lc}, there comes a point, actually a fold bifurcation of limit cycles, where the green unstable limit cycle collides with the red resonance captured limit cycle.
They annihilate one another leaving the blue limit cycle as the only stable attractor.
Consequently transience ensues as the dynamics is exponentially attracted to its new higher in terms of rotating speed but lower in terms of cart vibrations state.
Physically this explains the resonance release event (cf.\ Fig~\ref{fg:sweep}) whereby the rotor is now free to rotate at a frequency approximately consistent with its supplied torque.
Conversely, decreasing the supplied torque force from its value in Fig.~\ref{fg:3lc}, the blue resonance released limit cycle collides with the green unstable limit cycle.
They annihilate one another in a fold bifurcation of limit cycles leaving only the red resonance captured limit cycle.
This can also be seen in Figs~\ref{fg:sweep} and \ref{fg:contin} but the speed-down 
for the rotor is more modest as the blue and red limit cycles are much closer together at this bifurcation than the higher torque fold bifurcation.
It should be noted that the red limit cycle follows the normal operation of the limit cycle on a rigid base for low torque force.
We stress that the red limit cycle undergoes no change in its dynamical state, i.e.\ bifurcation, as it becomes captured by the resonance of the cart.
Instead, small oscillations always present in the cart and rotating rotor are merely amplified in the resonance zone.
As is often the case when a resonance curved is transversed, there is a phase difference of half the period between the red and blue limit cycles.
This can been seen in Fig.~\ref{fg:symm} as the displacement proceeds the velocity by a quarter period, the opposite is the case for the resonance released cycle.
Lastly we'd like to mention that the rate of change of the torque is important in the physical observation of the effect as the torque may be already significantly higher before 
the transient behaviour has had time to settle, this is well emphasised in Ref.~\cite{Evan1976}.

\section{Larger imbalance in the motor - Symmetry-broken limit cycles}\label{sec:larger}

\begin{figure}[ht!]
\center
\includegraphics[width=0.7\linewidth]{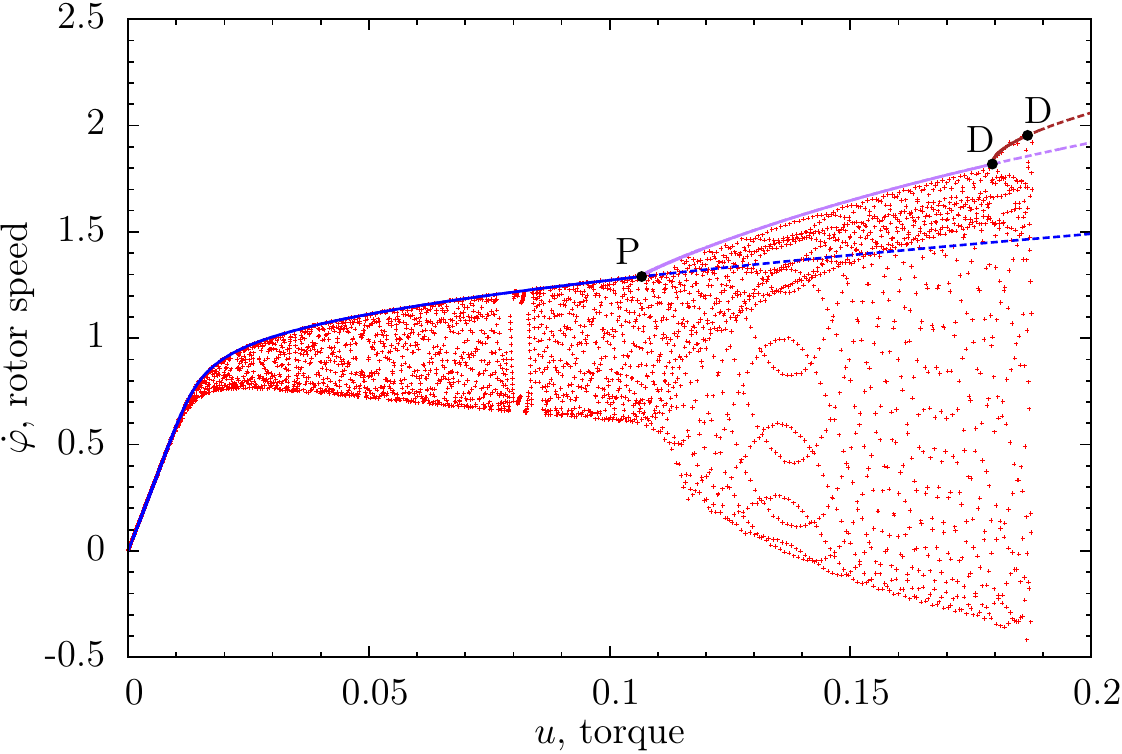}
\caption{\label{fg:PDD}
For $\epsilon=0.0125$, the torque is slowly increased  showing the resonance captured event. 
The dynamics undergo several bifurcations before at approximately $u=0.19$ 
the resonance captured dynamics is released.
Over plotted lines shows the maximum rotor speed of the limit cycles from continuation data.
Blue line follows a S-type cycle (cf.\ Fig.~\ref{fg:symm}), the purple line are M-type cycles (cf.\ Fig.~\ref{fg:sym_br}), and brown line shows a M-type cycle of doubled-period (cf.\ Fig~\ref{fg:sb-dp}). 
Solid lines are stable and dashed lines show unstable limit cycles. 
Bifurcations points are shown as solid black dots and labelled P to denote a supercritical pitchfork bifurcation of limit cycles and D to denote a period doubling bifurcation.}
\end{figure}

\begin{figure}[!bt]
\center
\includegraphics[width=0.8\linewidth]{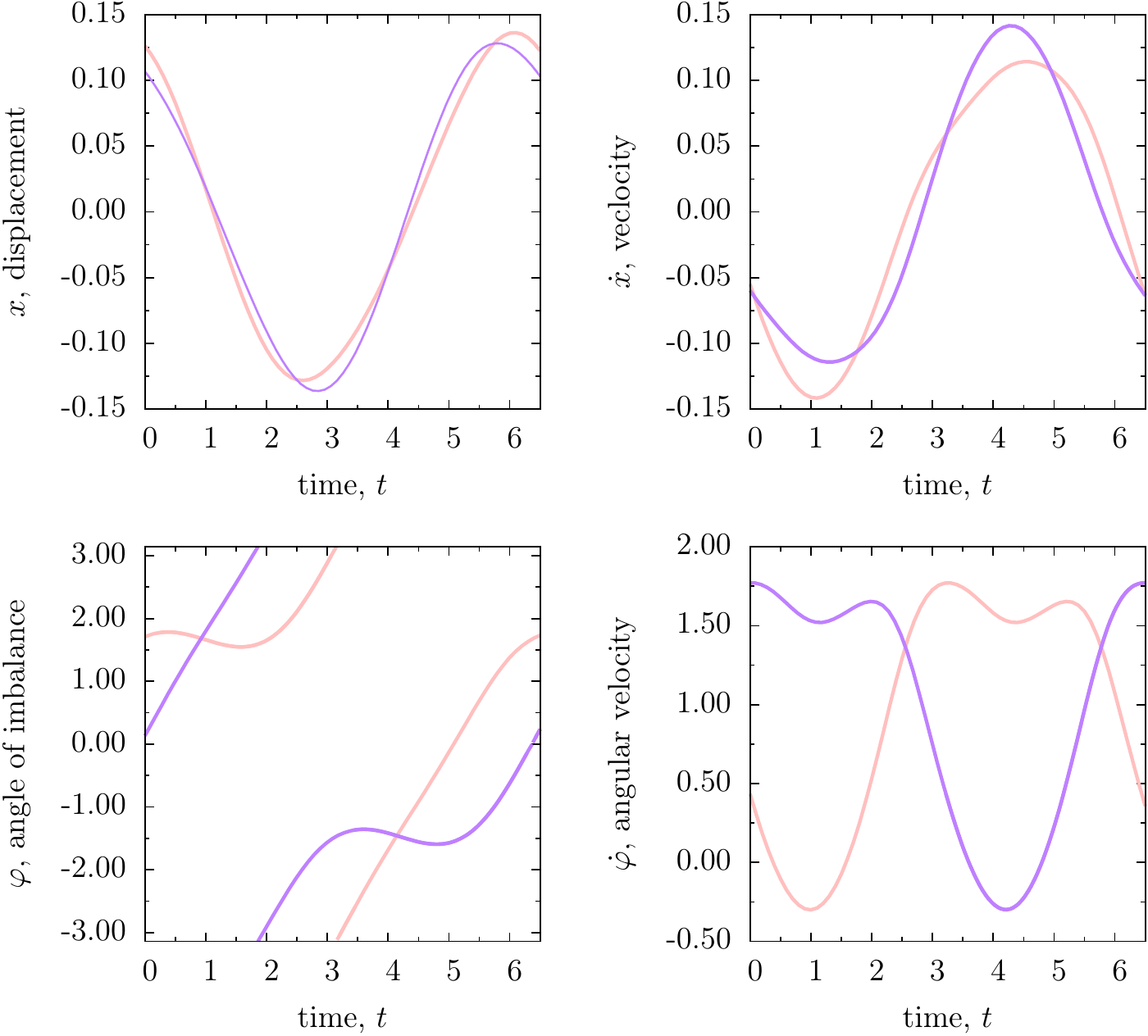}
\caption{\label{fg:sym_br} For $u=0.17$ and $\epsilon=0.0125$, time-traces showing a symmetry-broken with respect to 
symmetry~\eqref{eq:z2}, resonance captured M-type cycle~\cite{Nikolaev1995}, which momentary rotor reversal.
Period $\tau=\frac{2\pi}{avg(\dot{\varphi})}=\frac{6.28}{0.97}=6.48$. $x$ can be seen to proceed $\dot{x}$ by 
$\frac{\tau}{4}$.}
\end{figure}

As imbalance is needed in the motor to create the resonance capture and therefore the Sommerfeld effect, n\"{a}ively one may assume that an increase in the overall imbalance may lead to resonance capture to occur over a greater range of the supplied torque. 
In this section, it will be seen that this is not the case, but instead a different sequence of bifurcations than in Sec.~\ref{sec:basic} are possible whilst still maintaining the Sommerfeld effect phenomenon.
The red points in Fig.~\ref{fg:PDD} are rastored data from an integration in time of the equations of motion~\eqref{eq:orig} as the torque parameter is varied, but sufficiently slowly to allow the dynamics to settle to its steady state at each step.
In order to concentrate on the resonance captured event, the resonance released dynamics (the blue curve in Figs.~\ref{fg:sweep}, \ref{fg:3lc} and~\ref{fg:contin}) are not displayed in Fig.~\ref{fg:PDD}.
As before, when the torque force is increased, the dynamics is captured into resonance.
It can be seen that the rotor speed becomes enthralled to the resonance frequency of the cart but oscillates in $\dot{\varphi}$ with an amplitude that modestly grows as the torque increases.
After $u=0.1$, there comes a point where the dynamics significantly changes.
Continuation using  software AUTO~\cite{AUTO} reveals the S-type cycle (Fig.~\ref{eq:slc}), 
crucial to the basic Sommerfeld effect mechanism outlined in Sec.~\ref{sec:basic}, 
looses stability when a Floquet multiplier crosses the unit circle at real part one.
This is a supercritical pitchfork bifurcation which is labelled P in diagram Fig.~\ref{fg:PDD}.
After this point the S-type cycle although unstable continues in a similar bifurcation sequence discussed in the previous section.

\begin{figure}[!bt]
\center
\includegraphics[width=0.8\linewidth]{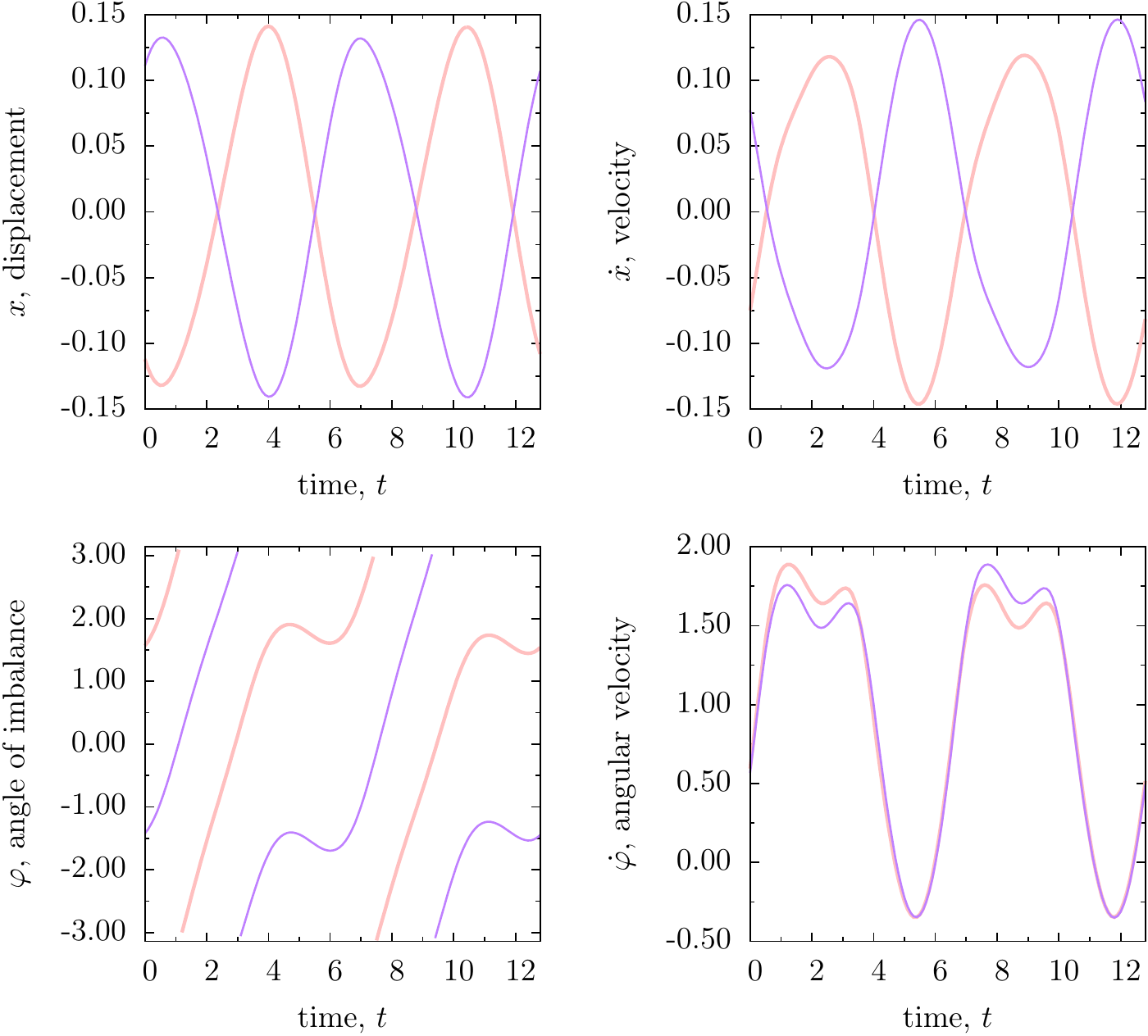}
\caption{\label{fg:sb-dp} For $u=0.18$ and $\epsilon=0.0125$, time-traces showing a symmetry-broken M-type limit cycle with respect to 
symmetry~\eqref{eq:z2}, with momentary rotor reversal.
Period has doubled $\tau=\frac{2\pi}{avg(\dot{\varphi})}\times 2=\frac{6.28}{0.98} \times 2 = 12.87$. $x$ can be seen 
to proceed $\dot{x}$ by $\frac{\tau}{4}$.}
\end{figure}

At the supercritical pitchfork bifurcation, stability is transferred to a not-previously-discussed type of limit cycle,
those that of course obey the symmetry~\eqref{eq:slc} but are no longer invariant to it and therefore are considered to be symmetry broken.
For these cycles applying the action of the symmetry~\eqref{eq:slc} results in a different albeit congruent cycle.
In the nomenclature of Ref.~\cite{Nikolaev1995}, these are called M-type cycles for ``mirror'' as they occur as a twin pair.
If a change of stability or local dynamics happens to one of them, it  must automatically happen to the other.
However as the symmetry is now broken, the symmetry~\eqref{eq:z2} no longer restricts the dynamics and the bifurcations that may occur individually.
As the torque force is increased further, it can be seen in Fig.~\ref{fg:PDD} that eventually the rotor speed oscillations go through zero and become negative.
Physically this would mean that the rotor, momentarily reverses to rotate in the clockwise direction before resuming its normal anti-clockwise revolutions.
Dynamically the point at which this occurs is not special, however we caution that for some coordinate systems, continuation of the M-type cycles may be difficult. 
A typical M-type cycle, displaying this reversal of rotor direction is shown in Fig.~\ref{fg:sym_br}.
In these time-traces, applying the action of the symmetry~\eqref{eq:z2} to the purple limit cycle results in the pink limit cycle and visa-versa.
The two M-type cycles may now be distinguished by the purple limit cycle having lower maximum positive displacement but higher maximum velocity than the pink limit cycle.
Both share the same angular velocity and therefore are difficult to distinguish in Fig.~\ref{fg:PDD}.
The next significant change in dynamics seen in Fig.~\ref{fg:PDD} is a period doubling bifurcation, labelled by the letter D.
We note that although S-type weren't permitted to undergo a period doubling, no such restriction occurs for M-type cycles.
At this point the twin cycles loose stability and a new M-type cycle pair with half the original frequency is born.
The displacement, velocity, angle of imbalance, and angular velocity of the rotor for this limit cycle is displayed in Fig.~\ref{fg:sb-dp}. 
The limit cycles are also symmetry-broken as can be seen from them having significantly different average displacement and velocity in the cart.
As before, applying symmetry~\ref{eq:z2} converts between the congruent purple and pink limit cycle pair. 
Shortly after the torque force is increased further, these period-doubled limit cycles in turn looses stability at another period-doubling bifurcation
creating new stable limit cycles.
Nevertheless, the stability of the resonance captured range is soon after lost and the dynamics is released from resonance to enter the higher rotor speed state.
Comparing Figs.~\ref{fg:sweep}, \ref{fg:contin} and~\ref{fg:PDD}, 
we note that this occurs at a significantly reduced torque strength than in Sec.~\ref{sec:basic},
meaning that the resonance captured region was reduced by the extra imbalance.

\section{Summary and conclusion}

In this manuscript, we directly articulate the bifurcation sequence that allows for the Sommerfeld effect,
with reference to the symmetries in the equations of motion.
A $\field{Z}_2$ phase space symmetry 
maintains a bistability of limit cycles over a larger range of the supplied torque force,
which facilitates the overall effect.
Furthermore the phase space symmetry is used to categorise the observed limit cycles 
and two parameter space symmetries are used to reduce the size of the parameter space needed to be considered.

\par

In the case of the basic mechanism, 
discussed in Sec.~\ref{sec:basic} and observed at low imbalance in the motor,
two fold bifurcations of limit cycles are created when a resonance curve is inclined back upon itself.
This results in a bistability with two stable limit cycle attractors separated by an unstable limit cycle.
The fold bifurcations allows for the asymmetry between increasing and decreasing the supplied torque to the motor.
Furthermore it explains the jump phenomena when the driving parameter passes outside the domain of existence of one of the stable limit cycles,
creating momentarily fast transience to the other stable limit cycle attractor.
In both cases, the unstable cycle annihilates a stable limit cycle at the bifurcation destroying the bistability.
At higher supplied torque, a greater difference in the relative kinetic and potential energies in the rotation and translation variables
leads to a more pronounced jump phenomenon.

\par

In Sec.~\ref{sec:larger}, the imbalance of the rotor is increased.
Via numerical continuation, the folded resonance curve of Sec.~\ref{sec:basic} is seen to occur, albeit it only affects unstable states. 
Instead a more complicated bifurcation sequence ensues and higher order multistabilities are observed as the limit cycle transfers stability at pitchfork bifurcation.
Invariance with respect to the $\field{Z}_2$ symmetry is broken permitting period-doubling bifurcations which subsequently occur.
These eventually lead to a loss of stability of the symmetry-broken states which significantly reduces the size of the resonance captured range and thus limits the Sommerfeld effect.
It may be thus hoped that a further study of the bifurcations along with manipulated symmetries may eliminate the phenomenon and permit a smooth passage through resonance.

%
%
\section*{Acknowledgements}

It is our great pleasure to present a paper at a conference with a session to honour Prof.\ Peter Hagedorn.
We thank him for his support of this research and for the many interesting discussions on many topics.

\par

This work was conducted during a two month visit (Oct./Nov. 2016) of the first author to the beautiful Pontifical Catholic University campus in Rio de Janeiro sponsored by FAPERJ-DFG exchange (HA 1060/58-1). He especially acknowledges the great hospitality of Prof.\ 
Sampaio's team there. The lead author is funded by the DFG (HA 1060/56-1).

%
%
\halfnormalsize
\bibliographystyle{acm} 	
\bibliography{dsta-2017} 	
%
%
%
\stamp{Eoin J Clerkin}{Ph.D.}{Technische Universit\"{a}t Darmstadt, fnb, Dynamics \& Vibrations Group}{Dolivostr. 15, 64293 Darmstadt, Germany}{eoin@clerkin.biz}{The author gave a presentation of this paper during one of the conference sessions.}
\stamp{Rubens Sampaio}{(professor)}{Pontif\'{i}cia Universidade Cat\'{o}lica do Rio de Janeiro, Departamento de Engenharia Mec\^{a}nica}{Mechanical Engineering Department, PUC-Rio, 22451-900 Rio de Janeiro, RJ, Brazil}{}{}
%
%
\end{document}